\documentclass[12pt]{article}
\usepackage{amssymb}
\usepackage{graphics}
\usepackage{epsfig}
\usepackage{a4wide}

\begin{document}
\newcommand{\eehww}{$e^+e^- \to H^0 W^+ W^-$ }
\newcommand{\eehwwg}{$e^+e^- \to H^0 W^+ W^-\gamma$ }
\newcommand{\eezh}{$e^+e^- \to Z^0H^0~$}

\title{ Precise predictions for the Higgs production
in association with a W-boson pair at ILC }
\author{ Song Mao, Ma Wen-Gan, Zhang Ren-You, Guo Lei, and Wang Shao-Ming \\
{\small Department of Modern Physics, University of Science and Technology}\\
{\small of China (USTC), Hefei, Anhui 230027, P.R.China}  }

\date{}
\maketitle \vskip 15mm
\begin{abstract}
The Higgs-boson production in association with a W-boson pair at
$e^+e^-$ linear colliders is one of the important processes in
probing the coupling between Higgs-boson and vector gauge bosons and
discovering the signature of new physics. We describe the impact of
the complete electroweak(EW) radiative corrections of ${\cal
O}(\alpha_{ew})$ to this process in the standard model(SM) at the
International Linear Collider(ILC), and investigate the dependence
of the lowest-order(LO) and EW next-to-leading order(NLO) corrected
cross sections on colliding energy $\sqrt{s}$ and Higg-boson mass.
The LO and NLO EW corrected distributions of the invariant mass of
W-boson pair and the transverse momenta of final $W$- and
Higgs-boson are presented. Our numerical results show that the
relative EW radiative correction($\delta_{ew}$) varies from
$-19.4\%$ to $0.2\%$ when $m_H=120~GeV$ and $\sqrt{s}$ goes up from
$300~GeV$ to $1.2~TeV$.
\end{abstract}

\vskip 2cm {\large{\bf Keywords}: Higgs-boson production, W-boson
pair,
electroweak interaction, radiative correction } \\

{\large\bf PACS: 12.15.Lk, 14.80.Bn, 14.70.Fm, 11.80.Fv }

\vfill \eject

\baselineskip=0.32in

\renewcommand{\theequation}{\arabic{section}.\arabic{equation}}
\renewcommand{\thesection}{\Roman{section}.}
\newcommand{\nb}{\nonumber}

\newcommand{\Dir}{\kern -6.4pt\Big{/}}
\newcommand{\Dirin}{\kern -10.4pt\Big{/}\kern 4.4pt}
\newcommand{\DDir}{\kern -7.6pt\Big{/}}
\newcommand{\DGir}{\kern -6.0pt\Big{/}}

\makeatletter      
\@addtoreset{equation}{section}
\makeatother       

\section{Introduction}
\par
The Higgs-boson plays a very important role in the standard
model(SM), it is responsible for the breaking of the
electroweak(EW) symmetry and the generation of masses for the
fundamental particles \cite{sm,higgs}. Unfortunately, it has not
been directly detected yet in experiments. Searching for
Higgs-boson within the SM and studying the phenomenology
concerning its properties are the important tasks for the present
and upcoming high energy colliders. Until now, we have obtained a
lot of experimental data which agree with the predictions of the
SM. These precise data constrain strongly the couplings of the
gauge-boson to fermions and the couplings between vector gauge
bosons, but present little information about the couplings between
Higgs-boson and gauge bosons. Previous LEP experiments have
provided the lower limit on the SM Higgs-boson mass as $114.4~
GeV$ at the $95\%$ confidence level, which is extracted from the
results of searches for $e^+e^- \to Z^0H^0$ production\cite{lower
mH,upper mH}. The present indirect evidences of the SM Higgs-boson
mass through EW precision measurements indicate the $95\%$ C.L.
upper bound as $m_H \lesssim 182~GeV$, when the lower limit on
$m_H$ is used in determination of this upper limit\cite{upper mH}.

\par
Compared to hadron machine, $e^+e^-$ linear collider has the
advantage of producing Higgs-boson in a particularly clean
environment. That is essential for studying Higgs-boson properties
since the cross section of Higgs-boson production is rather small
due to the fact that the scalar Higgs-boson couples mainly to
heavy particles. Furthermore, in the upper and lower limits for
the SM Higgs-boson mass, $H^0$ decays mainly into the bottom-quark
pair allowed by phase-space. Therefore, the properties of
Higgs-boson with the mass in present limitation range, are very
hard to be precisely probed at hadron colliders because of huge
QCD backgrounds.

\par
The future International Linear Collider (ILC) is designed as a
machine with the entire colliding energy in a range of $200~GeV
<\sqrt{s}<500~GeV$ and an integrated luminosity of around
$500~(fb)^{-1}$ in four years. The machine could be upgraded to
$\sqrt{s}\sim 1~TeV$ with an integrated luminosity of $1~(ab)^{-1}$
in three years\cite{ILC}. At this machine, Higgs-boson would be
produced mainly via the Higgs-boson strahlung process, and the
coupling of Higgs-boson to $Z$-bosons is probed best in the
measurement of the cross sections of the Higgs-boson strahlung
process $e^+e^-\to H^0Z^0$ and the $WW/ZZ$ fusion processes $e^+e^-
\to H^0 \nu \bar{\nu}$ and $e^+e^- \to H^0 e^+e^-$. The ILC would be
an ideal machine to observe and study in detail an intermediate-mass
or even a heavier Higgs-boson \cite{hamburg,saariselka}. It is sure
that once the neutral Higgs-boson is discovered and its mass is
determined, the $HVV$ production through $e^+e^- \to
H^0VV~(V=Z,W,\gamma)$ processes may provide the detail information
of the coupling between Higgs-boson and vector gauge bosons, which
directly reflects the role of the Higgs-boson in EW symmetry
breaking. Moreover, a theoretical accurate estimation of these
processes is essential, since $e^+e^- \to H^0VV~(V=Z,W,\gamma)$
processes could be potential backgrounds for possible new physics.
The investigation of the processes $e^+e^- \to H^0 W^+W^-$, $e^+e^-
\to H^0Z^0Z^0$ and $e^+e^- \to H^0Z^0 \gamma$ for testing the
couplings between Higgs-boson and gauge bosons were studied at the
tree-level in Ref.\cite{eehvv-1,eehvv}. The coupling of $g_{HZZ}$
can be determined at a few percent level for a $120~GeV$ Higgs-boson
with an integrated luminosity of $500~fb^{-1}$ from the production
cross section of the process $e^+e^- \to H^0Z^0Z^0$ as shown in
Ref.\cite{Precision}. Recently, Y.-J. Zhou et al., studied the full
${\cal O}(\alpha_{ew})$ EW corrections to the process $e^+e^- \to
H^0Z^0Z^0$\cite{eehzz}. They conclude that the ${\cal
O}(\alpha_{{ew}})$ EW radiative corrections to $e^+e^- \to
H^0Z^0Z^0$ process are in the range between $-15\%$ and $1.0\%$ when
$\sqrt{s}$ varies from $400~GeV$ to $2~TeV$, which should be taken
into consideration in the future precise experiments. The process of
Higgs-boson production in association with a pair of W-bosons at the
ILC is another significant process. It is not only important in
probing $g_{HWW}$ coupling, but also possible to provide further
tests for the quadrilinear couplings, such as $HWWZ$ and
$HWW\gamma$, which do not exist at tree-level in the SM, because
these quadrilinear couplings would induce deviations from the SM
predicted observables. That means the precise theoretical prediction
is necessary in experimental data analysis for \eehww process.

\par
In this work we calculate the full one-loop EW corrections to the
process \eehww in the SM. The paper is arranged as follows: In
Section II we give the calculation description of the Born cross
section, and the calculation of full ${\cal O}(\alpha_{ew}^4)$ EW
radiative contribution to the \eehww process is provided in section
III. In Section IV we present some numerical results and discussion,
and finally a short summary is given.

\vskip 10mm
\section{LO contributions of ${\cal O}(\alpha_{ew}^3)$ to \eehww process}
\par
\begin{figure*}
\begin{center}
\includegraphics*[120pt,400pt][540pt,710pt]{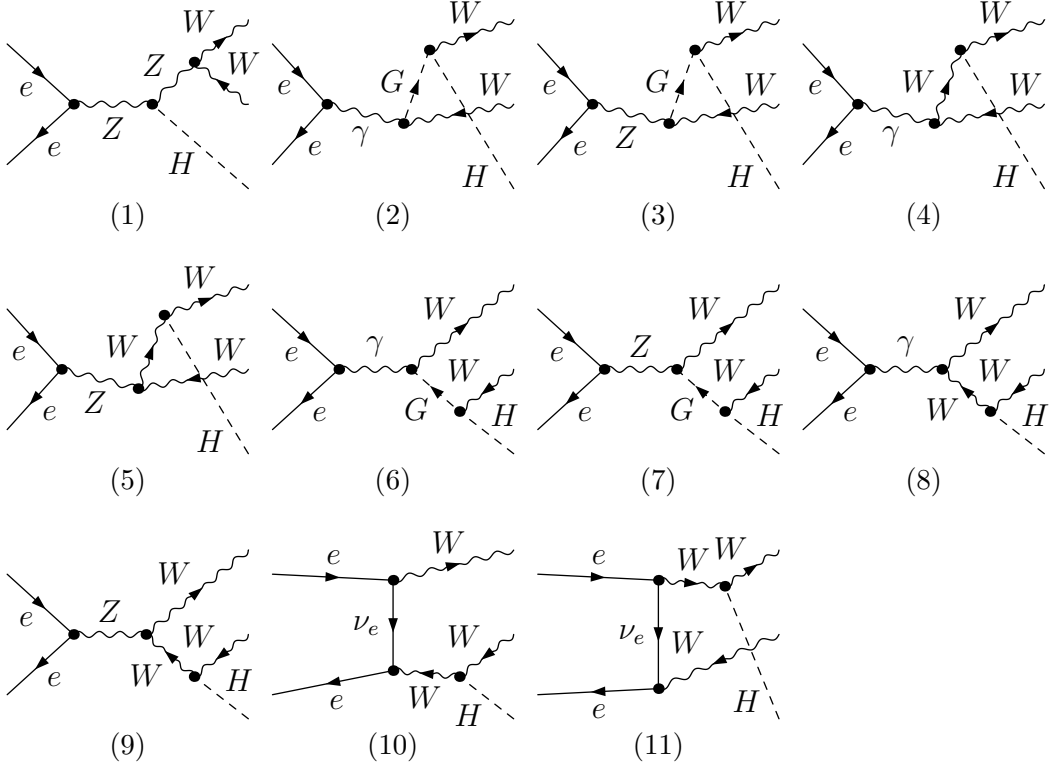}
\caption{\label{fig1} The full set of tree-level Feynman diagrams
for the \eehww process neglecting the electron-Higgs coupling. }
\end{center}
\end{figure*}

In our LO and NLO calculations we apply FeynArts3.3
package\cite{fey} to generate Feynman diagrams and convert them to
corresponding amplitudes. The amplitude reductions are mainly
implemented by applying FormCalc5.3 programs\cite{formloop}. We
use the 't Hooft-Feynman gauge in the leading-order calculation,
if there is no other statement. The contribution to the process
\eehww at the order of ${\cal O}(\alpha_{ew}^3)$ in the SM is
based on pure electroweak interactions. We ignore the contribution
from the Feynman diagrams which involve electron-Higgs Yukawa
coupling, because the concerning Yukawa coupling strength is
proportional to the electron mass, and the electron mass is
negligible comparing with the colliding energy $\sqrt{s}$ at the
ILC, Then there are eleven Feynman diagrams which we consider in
the calculation for the process \eehww at the tree-level(shown in
Fig.\ref{fig1}). The Feynman diagrams in Fig.\ref{fig1}
topologically can be divided into s-channel diagrams with
intermediate $\gamma$- or $Z^0$-boson, and t-channel diagrams with
$\nu_e$-exchanging. The notations for the process \eehww are
defined as
\begin{equation}
\label{process} e^+(p_1)+e^-(p_2) \to H^0(p_3)+ W^+(p_4)+W^-(p_5),
\end{equation}
where $p_i~(i=1-5)$ label the four-momenta of incoming positron,
electron and outgoing particles, respectively. The differential
cross section for the process \eehww at the tree-level is then
obtained as
\begin{eqnarray} \label{cross}
d\sigma_{tree} =\frac{1}{4} \sum_{spin}|{\cal M}_{tree}|^2 d\Phi_3 ,
\end{eqnarray}
where ${\cal M}_{tree}$ is the amplitude of all the tree-level
diagrams shown in Fig.\ref{fig1}. The factor $\frac{1}{4}$ is due to
taking average over the spins of the initial particles. $d\Phi_3$ is
the three-particle phase space element defined as
\begin{eqnarray}
d\Phi_3=\delta^{(4)} \left( p_1+p_2-\sum_{i=3}^5 p_i \right)
\prod_{j=3}^5 \frac{d^3 \textbf{\textsl{p}}_j}{(2 \pi)^3 2 E_j}.
\end{eqnarray}

\vskip 10mm
\section{Virtual and real photon emission corrections to \eehww process}
\par
\begin{figure}[htbp]
\vspace*{-0.3cm} \centering
\includegraphics*[130pt,210pt][540pt,710pt]{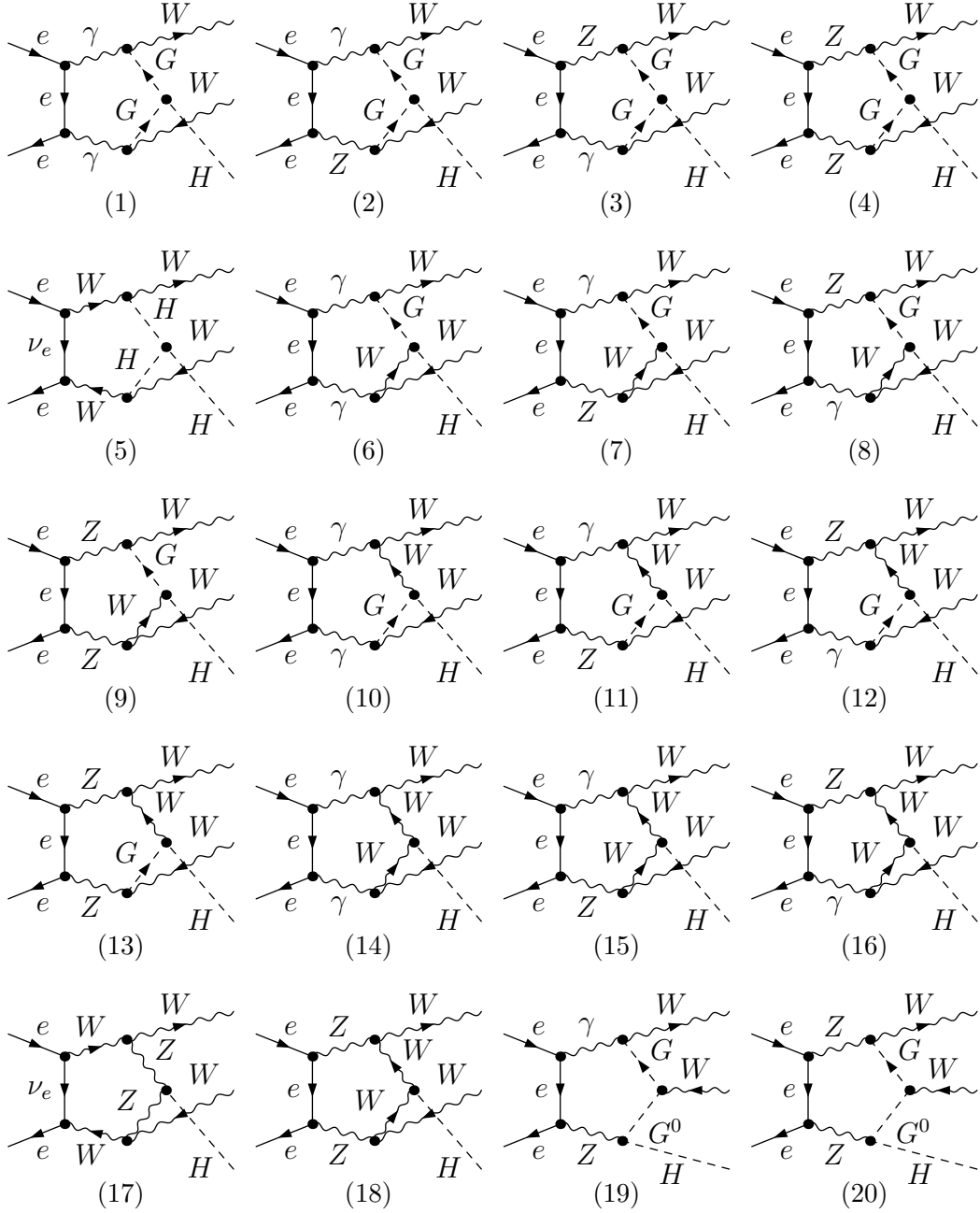}
\vspace*{-0.3cm} \centering \caption{\label{fig2} Some
representative pentagon Feynman diagrams for the process \eehww.}
\end{figure}

The one-loop Feynman diagrams and relevant counterterm diagrams of
process \eehww are also generated by using FeynArts3.3 package. At
the EW NLO, there are as many as 2770 one-loop Feynman diagrams
being taken into account in our calculation, and they can be
classified into self-energy, triangle(3-point), box(4-point),
pentagon(5-point) and counterterm diagrams. There are 82 pentagon
diagrams. A representative set of pentagon diagrams can be found
in Fig.\ref{fig2}. The calculation of the one-loop diagrams has
been performed in the conventional 't Hooft--Feynman gauge. The
virtual contribution of ${\cal O}(\alpha_{ew}^4)$ to \eehww
process can be expressed as\cite{hepdata}
\begin{eqnarray}
\Delta\sigma_{{\rm virtual}} = \sigma_{{\rm tree}} \delta_{{\rm
virtual}} = \frac{(2 \pi)^4}{2 |\vec{p}_1| \sqrt{s}} \int
\frac{1}{4}{\rm d} \Phi_3 \sum_{{\rm spin}} {\rm Re} \left( {\cal
M}_{{\rm tree}} {\cal M}_{{\rm virtual}}^{\dag} \right),
\end{eqnarray}
where $\vec{p}_1$ is the c.m.s. spatial momentum of the incoming
positron. ${\cal M}_{{\rm virtual}}$ represents the amplitude of
${\cal O}(\alpha_{ew}^4)$ order Feynman diagrams including EW
one-loop and counterterm diagrams. We use the dimensional
regularization(DR) method to isolate the ultraviolet(UV) and
infrared(IR) singularities. The numerical calculation of the one-,
two-, three- and four-point integral functions are implemented by
using the expressions presented in Refs.\cite{OneTwoThree,Four}.
The implementations of the scalar, vector and tensor five-point
integrals are done exactly by using the approach presented in
Refs.\cite{Passarino,Five}. The ${\cal O}(\alpha_{ew}^4)$
calculation of the cross section for hard photon radiation process
\eehwwg is accomplished by using Grace2.2.1 package\cite{Grace}.

\par
The total unrenormalized amplitude corresponding to all the
one-loop Feynman diagrams contains both ultraviolet (UV) and
infrared (IR) divergences. The UV and IR divergences in loop
integrals are isolated by adopting the dimensional
regularization(DR) scheme. The relevant fields are renormalized by
taking the on-mass-shell (OMS) scheme \cite{COMS scheme}. As we
expect, the UV divergence contributed by virtual one-loop diagrams
can be cancelled by that contributed from the counterterms exactly
both analytically and numerically.

\par
In our one-loop calculation there exists soft IR divergence, but no
collinear IR singularity, because we keep nonzero electron(positron)
mass. The soft IR divergence in virtual photonic corrections for the
process \eehww can be exactly cancelled by adding the real photonic
bremsstrahlung corrections to this process in the soft photon limit.
In the real photon emission process, denoted as
\begin{eqnarray}
\label{realphoton}
 e^+(p_1)+e^-(p_2) \to H^0(p_3)+W^+(p_4)+W^-(p_5)+\gamma(p_6),
\end{eqnarray}
a real photon radiates from the electron/positron or W-bosons, and
can be soft or hard. We adopt the general phase-space-slicing (PSS)
method \cite{PSS} to isolate the soft photon emission singularity
part in the real photon emission process \eehwwg. The bremsstrahlung
phase space is divided into singular and non-singular regions, and
the cross section of the \eehwwg process is decomposed into soft and
hard terms,
\begin{equation}
\Delta\sigma_{{real}}=\Delta\sigma_{{soft}}+\Delta\sigma_{{hard}}=
\sigma_{tree}(\delta_{{soft}}+\delta_{{hard}}).
\end{equation}
We can consider that the radiated photon in the reaction \eehwwg
can be either soft or hard, depending on the photon
energy($E_{6}$) in the center of mass system(c.m.s.) frame. The
criterion is like that: if $E_{6} \leq \Delta E$, the radiated
photon is soft, otherwise it is hard, where $\Delta E$ is defined
as $\Delta E\equiv\delta_sE_{b}$, and $E_b$ is the electron beam
energy in the c.m.s. frame and equals to $\sqrt{s}/2$. Then
theoretically both $\Delta\sigma_{{soft}}$ and
$\Delta\sigma_{{hard}}$ should depend on the arbitrary soft cutoff
$\delta_s$, but the total EW one-loop
correction($\Delta\sigma_{tot}$) and $\Delta\sigma_{{real}}$
should be cutoff $\delta_s$ independent. If the IR singularity in
the soft photon emission process is cancelled exactly with that
from the virtual photonic corrections, the independence of
$\Delta\sigma_{tot}(\equiv \Delta\sigma_{virtual}
+\Delta\sigma_{real})$ on the cutoff $\delta_s$ and fictitious
photon mass $m_{\gamma}$, could be demonstrated in our
calculculation. Since generally we take the soft cutoff $\delta_s$
to be a small value during our calculations, the terms of order
$\Delta E/E_{b}$ can be neglected and the soft contribution can be
evaluated by using the soft photon approximation analytically
\cite{COMS scheme, soft r approximation}
\begin{eqnarray}
\label{soft part} {\rm d} \Delta\sigma_{{\rm soft}} = -{\rm d}
\sigma_{tree} \frac{\alpha_{{\rm ew}}}{2 \pi^2}
 \int_{|\vec{p}_6| \leq \Delta E}\frac{{\rm d}^3 \vec{p}_6}{2 p_6^0} \left[
 \frac{p_1}{p_1\cdot p_6}-\frac{p_2}{p_2\cdot p_6}
 -\frac{p_4}{p_4\cdot p_6}+\frac{p_5}{p_5\cdot p_6} \right]^2,
\end{eqnarray}
where we define the four-momentum of radiated photon as
$p_6=(p_6^0,\vec{p}_6)$. As shown in Eq.(\ref{soft part}), the
soft contribution has an IR singularity at $\Delta E= 0$, which
can be cancelled exactly with that from the virtual photonic
corrections. The hard contribution is UV and IR finite, and can be
computed directly by using the Monte Carlo technique. Finally, the
corrected total cross section($\sigma_{tot}$) up to the order of
${\cal O}(\alpha^4_{ew})$ for the \eehww process is obtained by
summing the ${\cal O}(\alpha^3_{ew})$ Born cross
section($\sigma_{\rm tree}$), the ${\cal O}(\alpha^4_{ew})$
virtual cross section($\Delta\sigma_{\rm virtual}$), and the
${\cal O}(\alpha^4_{ew})$ cross section of the real photon
emission process \eehwwg($\Delta\sigma_{\rm real}$).
\begin{eqnarray}
\sigma_{{\rm tot}}=\sigma_{{\rm tree}} + \Delta\sigma_{{\rm tot}}=
\sigma_{{\rm tree}} + \Delta\sigma_{{\rm virtual}} +
\Delta\sigma_{{\rm real}} = \sigma_{{\rm tree}} \left( 1 +
\delta_{ew} \right),
\end{eqnarray}
where $\delta_{ew} = \delta_{{\rm virtual}} + \delta_{{\rm soft}}
+ \delta_{{\rm hard}}$ is the full ${\cal O}(\alpha_{ew})$
electroweak relative correction.

\vskip 10mm
\section{Numerical results and discussion}
\par
For the numerically verification of the gauge invariance in our LO
calculation, we used Grace2.2.1 and FeynArts3.3/FormCalc5.3
packages separately, and adopted 't Hooft-Feynman gauge and
unitary gauge respectively in calculating the Born cross section
of the process \eehww, and got the coincident numerical results.
We use also both Grace2.2.1 and FeynArts3.3/FormCalc5.3 packages
to calculate the tree-level cross section of process \eehww in 't
Hooft-Feynman gauge with the input parameters used in
Ref.\cite{eehvv}, and compare our numerical results with those in
Ref.\cite{eehvv} in order to verify again the correctness of our
tree-level calculation. The input parameters used in
Ref.\cite{eehvv} are taken as: $m_Z=91.18~GeV$, $m_W=80.1~GeV$,
$\alpha_{ew}(m_Z)=1/128$, $\sqrt{s}=500~GeV$ and an effective
$\sin^2 \theta_W=0.232$. The numerical results of the tree-level
cross section for the process \eehww are listed in Table
\ref{tab1}. We can see that Table \ref{tab1} yields results in
mutual agreement between ours and those presented in
Ref.\cite{eehvv}.
\begin{table}
\begin{center}
\begin{tabular}{|c|c|c|c|}
\hline $m_H (GeV)$ & $\sigma_{tree}(fb)$(Ref.\cite{eehvv})
& $\sigma_{tree}(fb)$(Grace) & $\sigma_{tree}(fb)$ (FeynArts) \\
\hline 120  & 5.63 & 5.6378(6)  & 5.6374(6)  \\ \hline 150  & 3.51 &
3.5123(3) & 3.5126(3) \\ \hline
200  & 1.47 & 1.4714(1) & 1.4715(1) \\
\hline
\end{tabular}
\end{center}
\begin{center}
\begin{minipage}{15cm}
\caption{\label{tab1} The comparison of the numerical results of the
tree-level cross sections for the process $e^+e^- \to H^0W^+W^-$
with the corresponding ones in Ref.\cite{eehvv}, by taking the same
input parameters as in Ref.\cite{eehvv} and using Grace2.2.1 and
FeynArts3.3/FormCalc5.3 packages separately. }
\end{minipage}
\end{center}
\end{table}

\par
The following numerical computation is performed in the
$\alpha_{ew}$-scheme and the initial EW input physical parameters
are taken as \cite{hepdata}:
\begin{eqnarray} \label{input1}
m_W&=&80.403~{\rm GeV},~~m_Z~=~91.1876~{\rm GeV},~~\alpha_{ew}(0)
= 1/137.03599911.
\end{eqnarray}
The charged leptons have the mass values as
\begin{eqnarray} \label{input2}
m_e&=&0.51099892~{\rm MeV},~m_\mu~=~105.658369~{\rm
MeV},~m_\tau~=~1776.99~{\rm MeV}.
\end{eqnarray}
For the quark masses, beside the top-quark mass $m_t=~172.5~{\rm
GeV}$, we take the set $m_u=m_d=66~{\rm MeV}$, $m_s~=~150~{\rm
MeV}$, $m_c~=~1.5~{\rm GeV}$, and $m_b~=~4.7~{\rm GeV}$. Since
quark-mixing effects are suppressed, we set the CKM matrix to the
unit matrix.

\par
Except the input parameters shown above, we have to fix the values
of the IR regulator $m_{\gamma}$, the fictitious photon mass, and
soft cutoff $\delta_s=\Delta E/E_b$ during our one-loop numerical
calculation. In fact, if the one-loop calculation is correct and
the IR divergency is really cancelled, the total cross section
should be independent of these two parameters. Our numerical
results show that the cross section contribution at ${\cal O}
(\alpha_{ew}^4)$ order,
$\Delta\sigma_{tot}=\Delta\sigma_{real}+\Delta\sigma_{virtual}$,
is invariant within the calculation errors when the fictitious
photon mass $m_{\gamma}$ varies from $10^{-15}~ GeV$ to
$10^{-1}~GeV$ in conditions of $\delta_s=10^{-3}$, $m_H=120~GeV$
and $\sqrt{s}=500~GeV$.

\par
\begin{figure}
\includegraphics[scale=0.4]{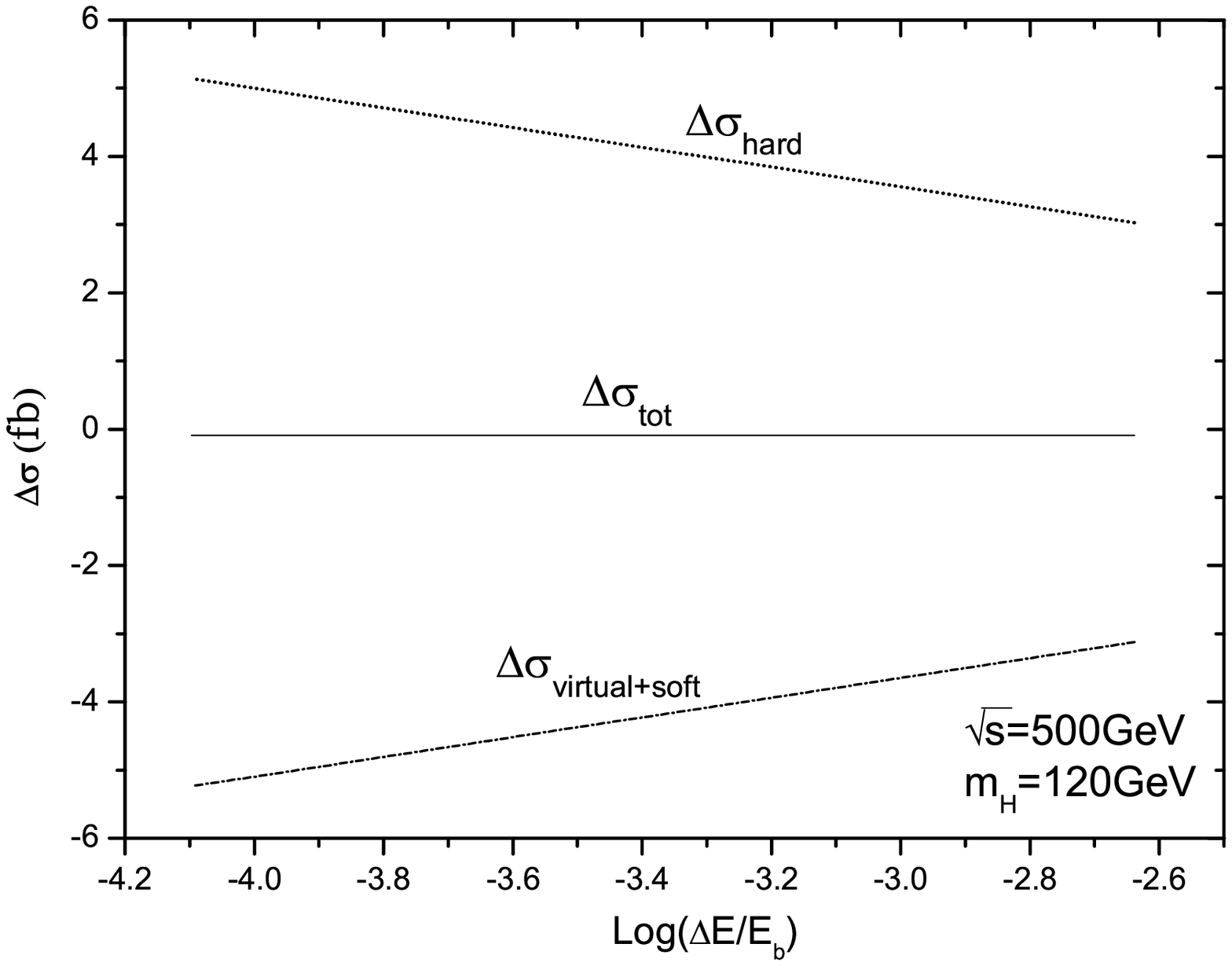}
\includegraphics[scale=0.4]{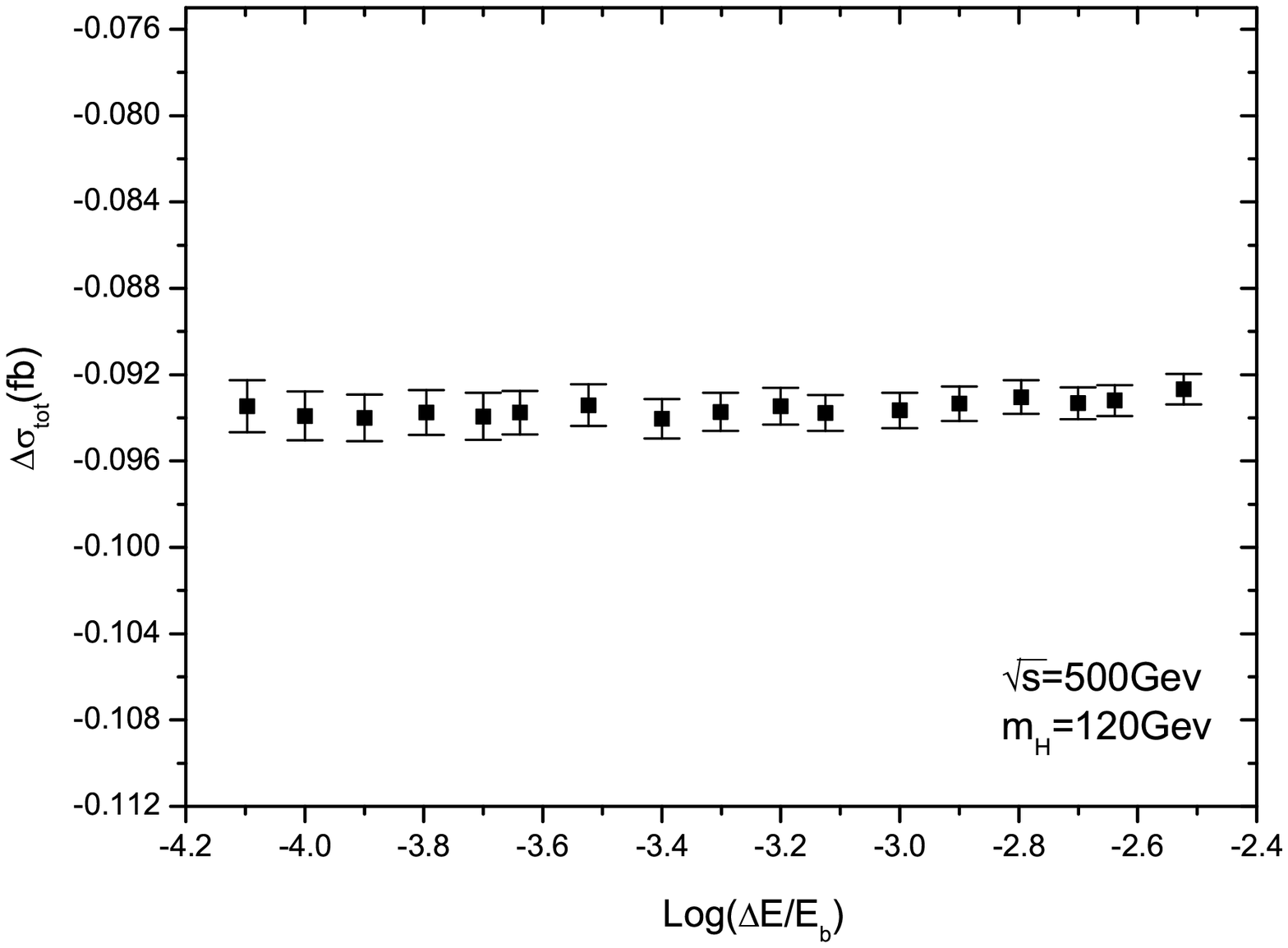}
\caption{\label{fig3} (a) The ${\cal O}(\alpha_{{\rm ew}}^4)$
contribution parts of cross section for \eehww process as the
functions of the soft cutoff $\delta_s\equiv \Delta E/E_b$ in
conditions of $m_{\gamma}=10^{-10}~GeV$, $m_H=120~GeV$ and
$\sqrt{s}=500~GeV$. (b) The amplified curve for
$\Delta\sigma_{tot}$ of Fig.3(a) versus $\delta_s$ including
calculation errors.}
\end{figure}

As a check of the correctness in our calculation of radiative
correction, we present Figs.\ref{fig3}(a-b) to show the
independence of the total ${\cal O}(\alpha_{{\rm ew}}^4)$
electroweak contribution to \eehww process on soft cutoff
$\delta_s$, assuming $m_{\gamma}=10^{-5}~GeV$, $m_{H} = 120~GeV$
and $\sqrt{s} = 500~GeV$. As shown in Fig.\ref{fig3}(a), both
$\Delta\sigma_{{\rm virtual}}+\Delta\sigma_{{\rm soft}}$ and
$\Delta\sigma_{{\rm hard}}$ strongly depend on soft cutoff
$\delta_s$, but the total ${\cal O}(\alpha_{{\rm ew}}^4)$ EW
relative radiative contribution
$\Delta\sigma_{tot}=\Delta\sigma_{{\rm
virtual}}+\Delta\sigma_{{\rm real}}$ is cutoff $\delta_s$
independent within the range of calculation errors as expected. In
order to show the curve of $\Delta\sigma_{tot}$ more clearly, we
present the amplified curve for $\Delta\sigma_{tot}$ including
calculation errors in Fig.\ref{fig3}(b). In further calculations,
we fix $m_{\gamma}=10^{-5}~GeV$ and $\delta_s=10^{-3}$.

\par
\begin{figure}
\centering
\includegraphics[scale=0.35]{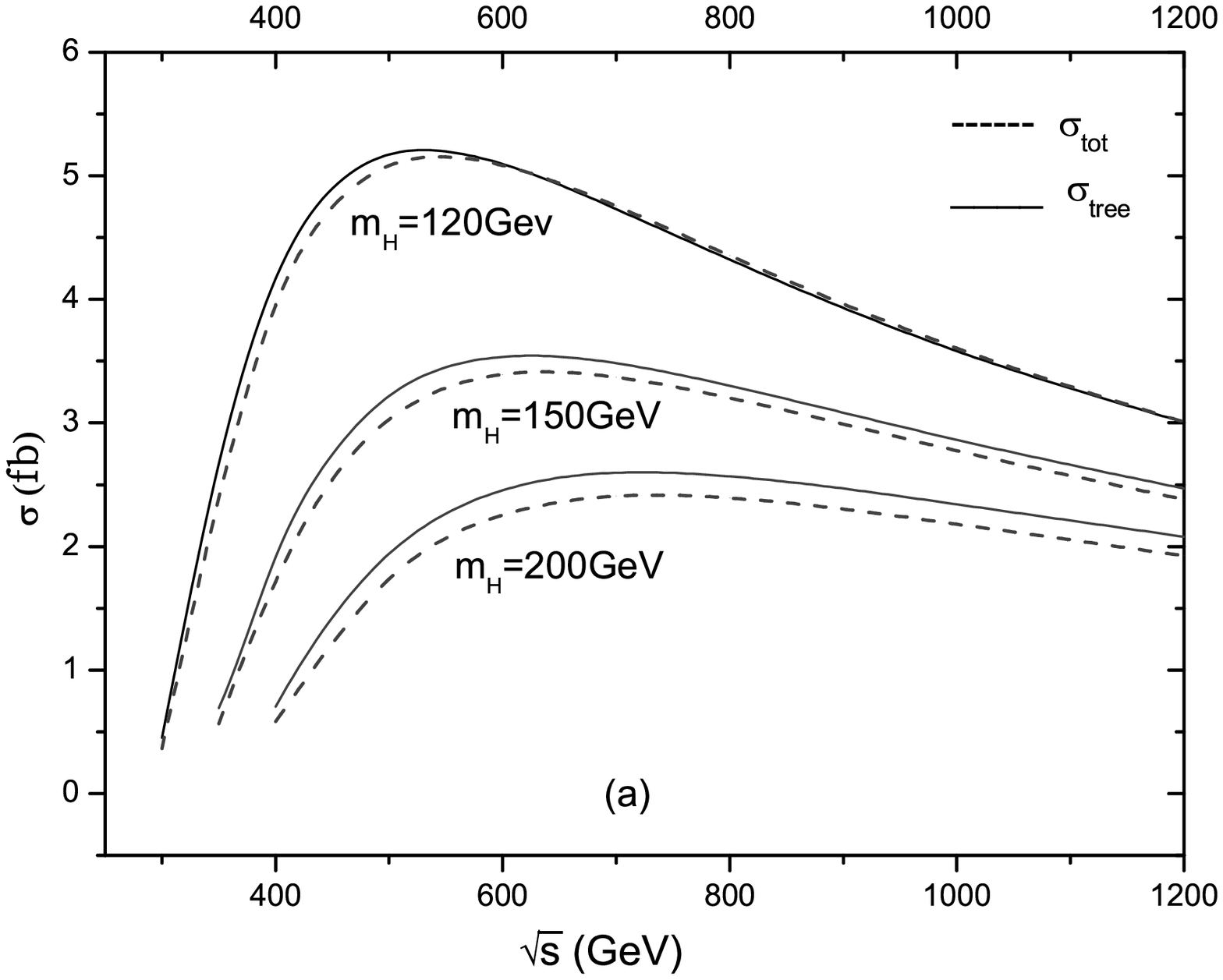}
\includegraphics[scale=0.35]{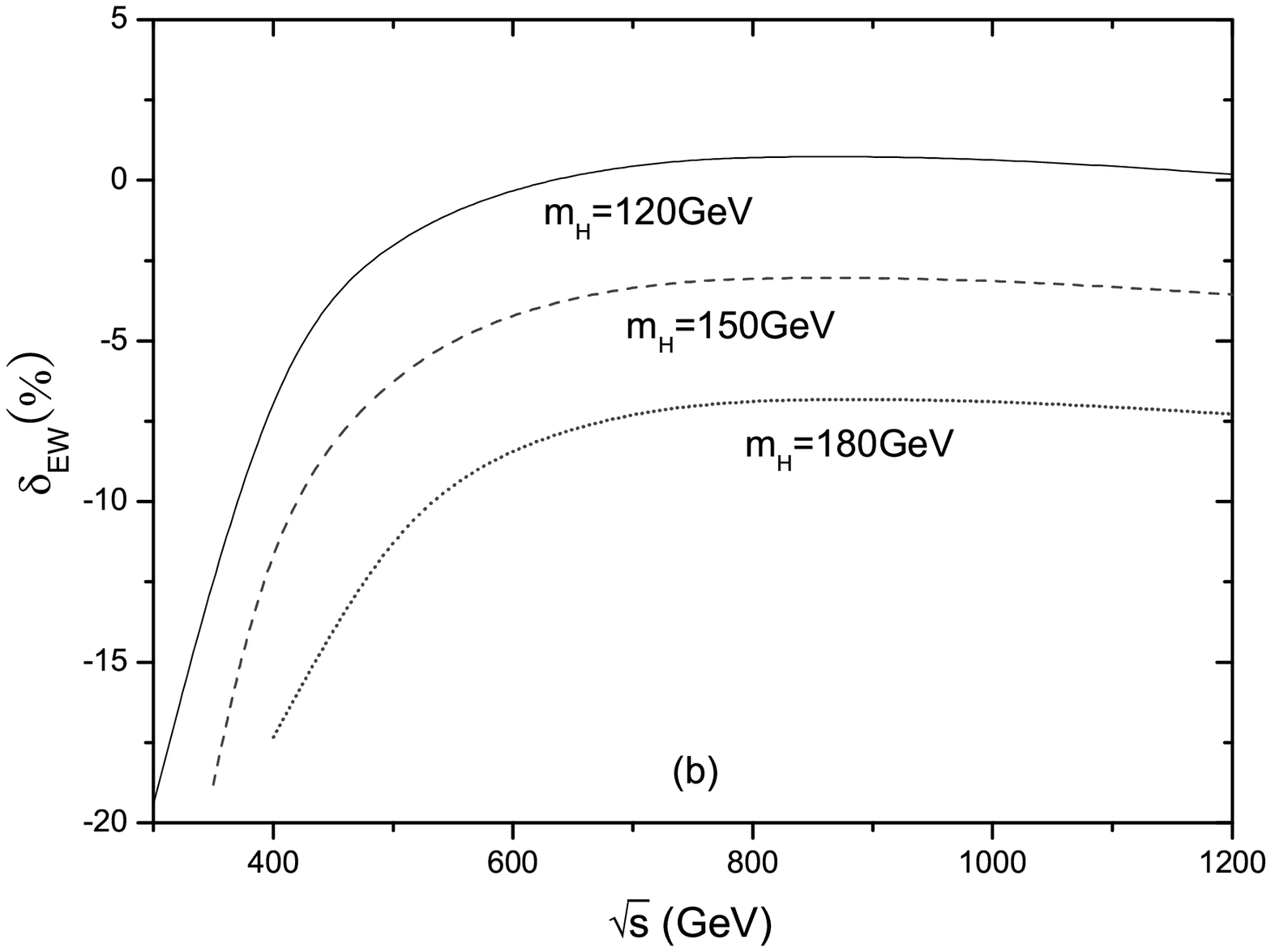}
\caption{\label{fig4} (a) The LO and EW one-loop corrected cross
sections for the process \eehww as the functions of colliding
energy $\sqrt{s}$ with $m_H=120~GeV$, $150~GeV$, $180~GeV$
respectively. (b) The corresponding relative EW radiative
corrections ($\delta_{ew}$) versus $\sqrt{s}$.}
\end{figure}

In Fig.\ref{fig4}(a) we present the curves of the LO and one-loop
EW corrected cross sections as the functions of colliding energy
$\sqrt{s}$, with $m_H=120~GeV$, $150~GeV$, $200~GeV$, separately.
Fig.\ref{fig4}(b) shows their corresponding relative EW radiative
corrections($\delta_{ew}\equiv \frac{
\Delta\sigma_{tot}}{\sigma_{tree}}$) for the data presented in
Fig.\ref{fig4}(a). We find from Figs.\ref{fig4}(a-b) that the LO
cross sections are sensitive to the Higgs-boson mass and generally
suppressed by the one-loop EW corrections except for $m_H=120~GeV$
and the colliding energy being in the range of $\sqrt{s} >
630~GeV$. Fig.\ref{fig4}(b) shows that the relative radiative
correction due to full EW one-loop contributions in the $\sqrt{s}$
vicinity close to the threshold of $H^0W^+W^-$ production, becomes
rather large. That is because of the Coulomb singularity effect
from the diagrams involving the instantaneous virtual photon
exchange in loop which has a small spatial momentum. To show the
numerical results as presented in Figs.\ref{fig4}(a-b) more
precisely, we list some typical numerical results of the
tree-level, one-loop EW corrected cross sections and the relative
EW radiative correction($\delta_{ew}\equiv
\Delta\sigma_{tot}/\sigma_{tree}$) for the process \eehww in Table
\ref{tab2}. There they are in conditions of $\sqrt{s}=400~GeV$,
$600~GeV$, $800~GeV$, $1000~GeV$, $1200~GeV$ and $m_H=120~GeV$,
$150~GeV$, $180~GeV$ separately. We can read out also from
Fig.\ref{fig4}(b) that when $m_H=120~GeV$ and $\sqrt{s}$ goes up
from $300~GeV$ to $1.2~TeV$, the relative EW radiative correction,
$\delta_{ew}$, increases from $-19.4\%$ to $0.2\%$.

\begin{table}
\begin{center}
\begin{tabular}{|c|c|c|c|c|}
\hline $\sqrt{s}$ (GeV) & $m_H (GeV)$ & $\sigma_{tree}(fb)$
& $\sigma_{tot}(fb)$ & $\delta_{ew}($\%$)$  \\
\hline  & 120  & 4.1676(4)  & 3.953(2)  & -5.15(5)  \\
   400  & 150  & 1.9144(2)  & 1.719(1)  & -10.21(5) \\
        & 180  & 0.70438(7) & 0.5823(5) & -17.33(7) \\
\hline  & 120  & 5.0945(5)  & 5.085(3)  & -0.19(6)  \\
   600  & 150  & 3.5357(3)  & 3.392(2)  & -4.06(6)  \\
        & 180  & 2.4562(3)  & 2.254(2)  & -8.23(8)  \\
\hline  & 120  & 4.3205(6)  & 4.354(3)  & 0.78(7)  \\
   800  & 150  & 3.3009(4)  & 3.202(2)  & -3.00(6)  \\
        & 180  & 2.5672(3)  & 2.393(2)  & -6.79(8)  \\
\hline   & 120 & 3.5817(6)  & 3.607(3)  & 0.71(8)  \\
   1000  & 150 & 2.8635(4)  & 2.776(2)  & -3.06(7)  \\
         & 180 & 2.3418(3)  & 2.182(2)  & -6.82(8) \\
\hline  & 120  & 3.0044(5)  & 3.010(3)  & 0.19(8)  \\
  1200  & 150  & 2.4678(4)  & 2.380(2)  & -3.56(8)  \\
        & 180  & 2.0770(3)  & 1.926(2)  & -7.27(9) \\
\hline
\end{tabular}
\end{center}
\begin{center}
\begin{minipage}{15cm}
\caption{\label{tab2} The numerical results of the tree-level,
one-loop EW corrected cross sections and the relative EW radiative
correction($\delta_{ew}\equiv \Delta\sigma_{tot}/\sigma_{tree}$)
for the process \eehww, by taking $\sqrt{s}=400~GeV$, $600~GeV$,
$800~GeV$, $1000~GeV$, $1200~GeV$ and $m_H=120~GeV$, $150~GeV$,
$180~GeV$ separately. }
\end{minipage}
\end{center}
\end{table}

\par
\begin{figure}
\centering
\includegraphics[scale=0.35]{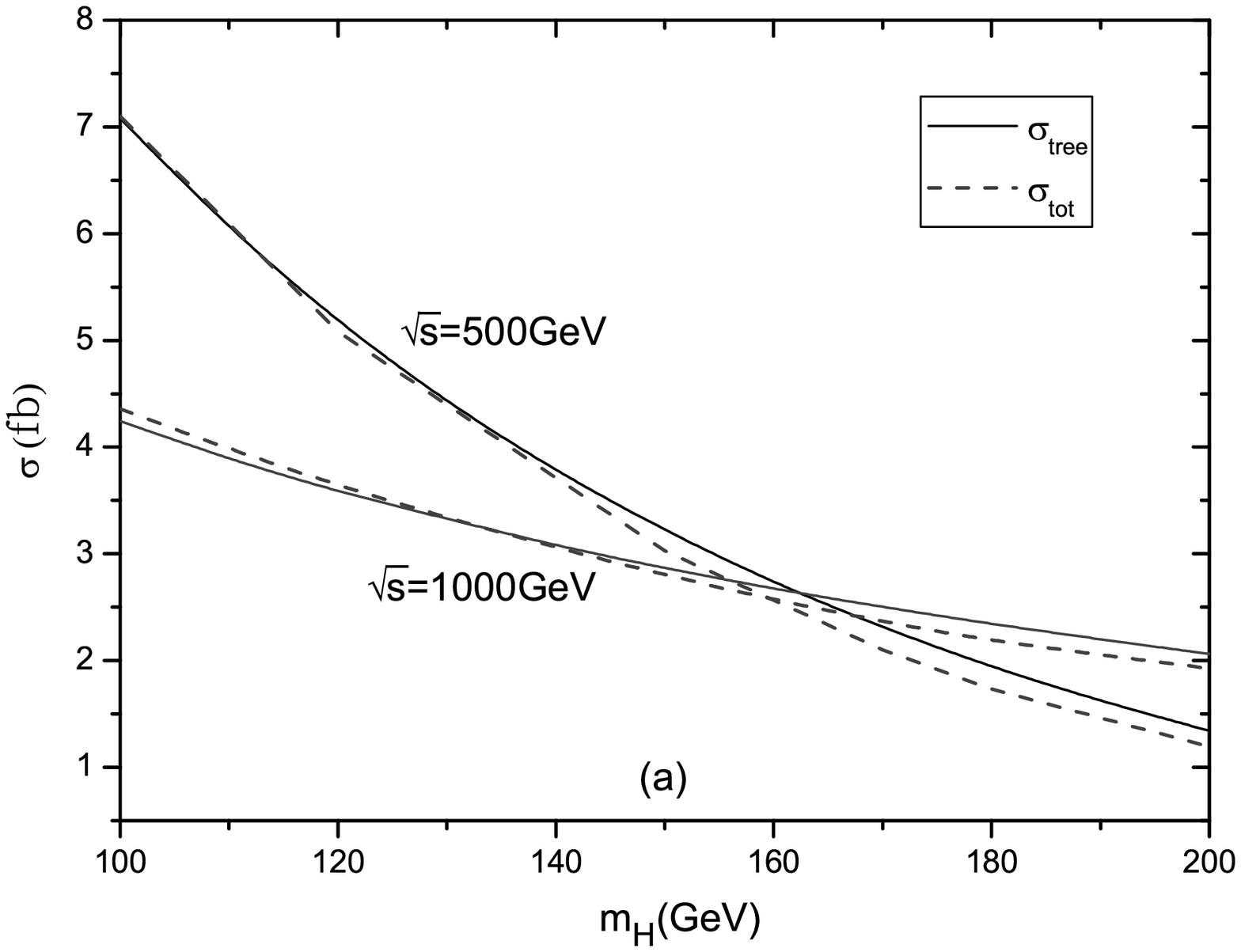}
\includegraphics[scale=0.35]{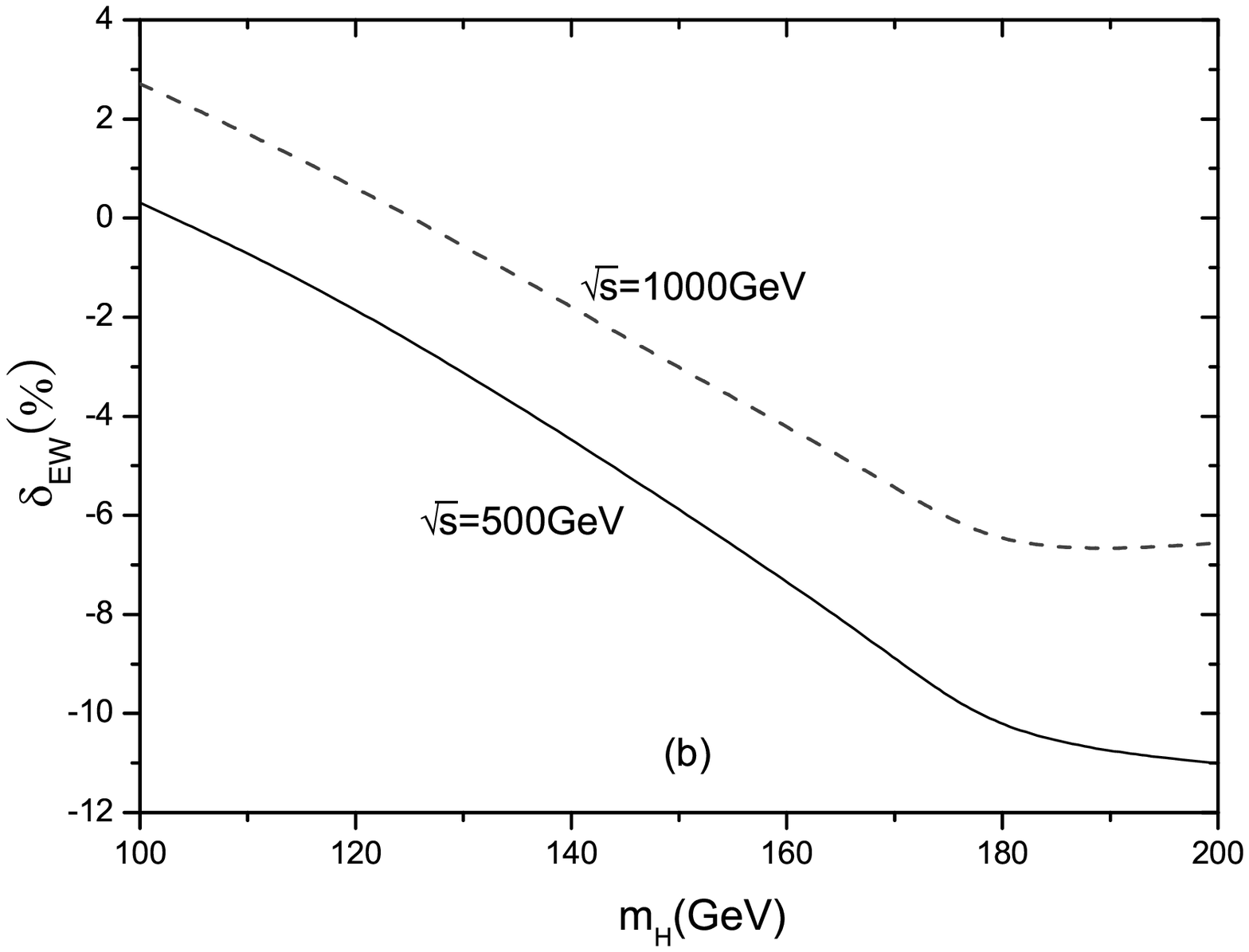}
\caption{\label{fig5} (a) The LO and EW one-loop corrected cross
sections for the process \eehww as the functions of the
Higgs-boson mass $m_H$ with $\sqrt{s}=500~GeV,~1000~GeV$. (b) The
corresponding relative EW radiative corrections versus Higgs-boson
mass. }
\end{figure}

In Fig.\ref{fig5}(a) we present the plot of the LO and EW one-loop
corrected cross sections for the process \eehww as the functions
of Higgs-boson mass $m_H$ by taking $\sqrt{s}=500~GeV$ and
$1000~GeV$ separately, and their corresponding relative EW
radiative corrections versus Higgs-boson mass are depicted in
Fig.\ref{fig5}(b). From these two figures we can see again the
cross sections at LO and up to one-loop order for the process
\eehww are all sensitive to the Higgs-boson mass as already shown
in Figs.\ref{fig4}(a-b). And both curves for $\sqrt{s}=500~GeV$
and $1000~GeV$ in Fig.\ref{fig5}(b) demonstrate that the absolute
relative EW radiative corrections for $\sqrt{s}=500~GeV$ and
$1000~GeV$ go up quickly, when Higgs-boson mass goes up from from
$100~GeV$ to $180~GeV$, but become relative stable when $m_H$ runs
from $180~GeV$ to $200~GeV$.

\par
Since we consider the CP-conserving SM, the distributions of
transverse momenta of $W^-$-boson should be the same as that of
$p_T^{W^+}$. Therefore, we shall not provide the distribution of
$p_T^{W^-}$ but only the $p_T^{W^+}$. We depict the differential
cross sections of transverse momentum of $H^0$-boson at the LO and
up to NLO ($d\sigma_{LO,NLO}/dp_T^{H^0}$) in Fig.\ref{fig5}(a),
and the distributions of $d\sigma_{LO}/dp_T^{W^+}$ and
$d\sigma_{NLO}/dp_T^{W^+}$in Fig.\ref{fig5}(b) separately, in the
conditions of $m_H=120~GeV$ and $\sqrt{s}=500~GeV$. We can see
from Figs.\ref{fig5}(a-b) that when the transverse momentum
$p_T^{H^0}$($p_T^{W^+}$) is in the range smaller than $90~GeV$,
the LO differential cross section of
$d\sigma_{LO}/dp_T^{H^0}$($d\sigma_{LO}/dp_T^{W^+}$) is slightly
enhanced by the EW one-loop corrections, while it is suppressed
when $p_T^{H^0}$($p_T^{W^+}$) is larger than $100~GeV$.
\begin{figure}
\centering
\includegraphics[scale=0.35]{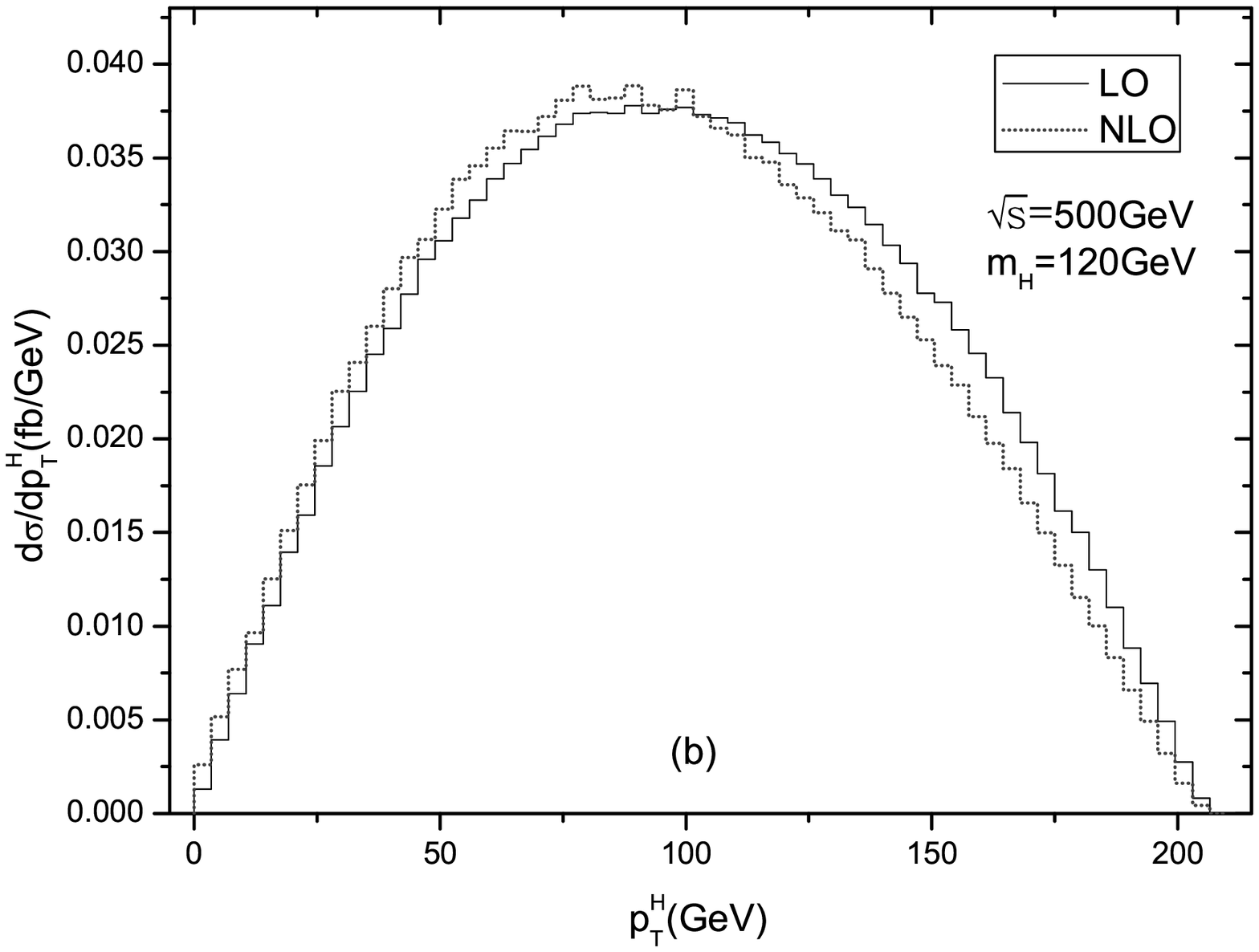}
\includegraphics[scale=0.35]{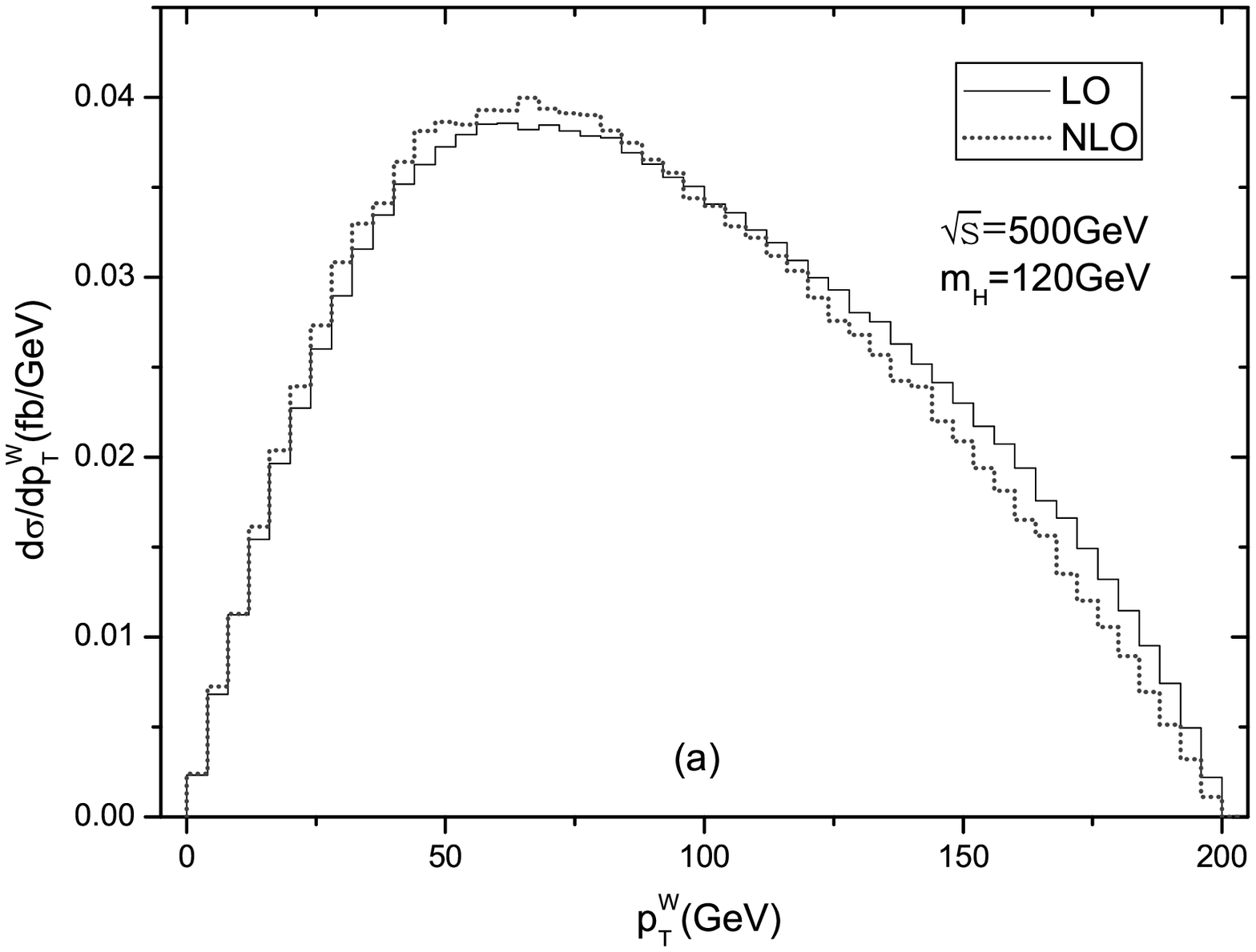}
\caption{\label{fig6} The distributions of the transverse momenta of
$H^0$- and $W^+$-boson for the \eehww process at the LO and up to EW
one-loop level with $\sqrt{s}=500~GeV$ and $m_H=120~GeV$. (a) for
$H^0$-boson, (b) for $W^+$-boson. }
\end{figure}

\par
\begin{figure}
\centering
\includegraphics[scale=0.4]{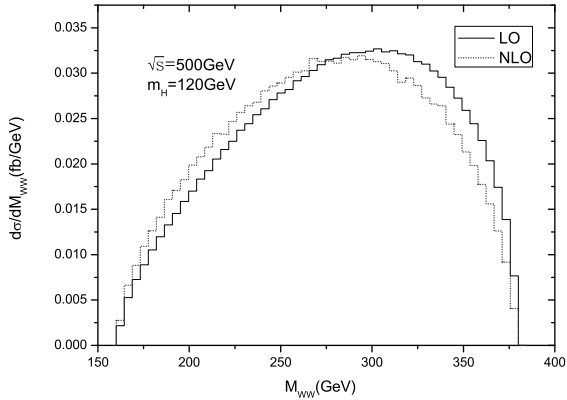}
\caption{\label{fig7} The distributions of the invariant mass of
W-pair at the LO and up to EW one-loop level when $m_H=120~GeV$
and $\sqrt{s}=500~GeV$. }
\end{figure}

We plot the distributions of W-pair invariant mass, denoted as
$M_{WW}$, at the LO and up to EW one-loop level in Fig.\ref{fig7}
by taking $m_H=120~GeV$ and $\sqrt{s}=500~GeV$. We can see from
the figure that when $M_{WW}<270~GeV$ the one-loop EW corrections
enhance the LO differential cross section $d\sigma_{LO}/dM_{WW}$
slightly, while in the $M_{WW}$ region being larger than $280~GeV$
the $d\sigma_{LO}/dM_{WW}$ is obviously suppressed by the EW
corrections.

\vskip 10mm
\section{Summary}
\par
The future $e^+e^-$ linear collider, International Linear
Collider(ILC), will probably offer the cleanest environment to
probe the SM more precisely and discover the signature of new
physics. In this paper we have shown that the phenomenological
effects due to the contribution from complete one-loop EW terms of
${\cal O}(\alpha_{ew}^4)$, can be demonstrated in the study of the
electron-positron into Higgs-boson in association with a W-boson
pair in the standard model(SM) for all energies ranging from
$300~GeV$ to $1.2~TeV$ at ILC. We discuss the dependence of the
effects coming from the EW NLO contribution to the cross section
of process \eehww, on colliding energy $\sqrt{s}$ and Higgs-boson
mass. We present the LO and EW NLO corrected distributions of the
transverse momenta of final particles and the differential cross
section of $W$-pair invariant mass. We find that the LO cross
section of process \eehww is sensitive to Higgs-boson mass and
generally suppressed by the one-loop EW corrections except when
$m_H=120~GeV$ and $\sqrt{s}
> 630~GeV$. Our numerical results show that when $m_H=120~GeV$ and
$\sqrt{s}$ goes up from $300~GeV$ to $1.2~TeV$ the relative EW
radiative correction varies from $-19.4\%$ to $0.2\%$. It shows
that the EW one-loop radiative  corrections to \eehww process can
be significant and should be included in any reliable analysis.

\vskip 10mm
\par
\noindent{\large\bf Acknowledgments:} This work was supported in
part by the National Natural Science Foundation of China,
Specialized Research Fund for the Doctoral Program of Higher
Education(SRFDP) and a special fund sponsored by Chinese Academy
of Sciences.

\end{document}